\documentclass[fleqn]{2020SCGE}
\usepackage{color}
\usepackage{float}
\usepackage{gensymb}
\usepackage{url}
\usepackage{hyperref}

\begin{document}

\ensubject{subject}


\ArticleType{Article}
\SpecialTopic{SPECIAL TOPIC: }
\Year{2020}
\Month{March}
\Vol{0}
\No{0}
\DOI{}
\ArtNo{000000}
\ReceiveDate{x}
\AcceptDate{x}

\title{FAST discovery of an extremely radio-faint millisecond pulsar from the \textit{Fermi}-LAT unassociated source 3FGL J0318.1+0252}{FAST discovery of an extremely radio-faint millisecond pulsar from the \textit{Fermi}-LAT unassociated source 3FGL J0318.1+0252}

\author[1]{\\Pei Wang$^{\dag}$}{}
\author[1,2]{Di Li$^{\ddag}$}{}
\author[3,18]{Colin J. Clark}{}
\author[4,5]{Pablo M. Saz Parkinson}{}
\author[6,7]{Xian Hou}{}
\author[1]{Weiwei Zhu}{}
\author[1]{Lei Qian}{}
\author[1]{\\Youling Yue}{}
\author[1]{Zhichen Pan}{}
\author[8]{Zhijie Liu}{}
\author[8]{Xuhong Yu}{}
\author[8]{Shanping You}{}
\author[8]{Xiaoyao Xie}{}
\author[9]{Qijun Zhi}{}
\author[8]{\\Hui Zhang}{}
\author[1]{Jumei Yao}{}
\author[1]{Jun Yan}{}
\author[1]{Chengmin Zhang}{}
\author[4,17]{Kwok Lung Fan}{}
\author[14]{Paul S. Ray}{}
\author[14]{Matthew Kerr}{}
\author[13]{\\David A. Smith}{}
\author[10]{Peter F. Michelson}{}
\author[11]{Elizabeth C. Ferrara}{}
\author[12]{David J. Thompson}{}
\author[15]{Zhiqiang Shen}{}
\author[16]{\\Na Wang}{}
\author[]{FAST $\&$ \textit{Fermi}-LAT Collaboration}{}
\footnote{$^{\dag}$ wangpei@nao.cas.cn}
\footnote{$^{\ddag}$ dili@nao.cas.cn}


\AuthorMark{Pei Wang}

\AuthorCitation{Pei Wang, Di Li, Colin J. Clark, Pablo Saz Parkinson, Xian Hou, Weiwei Zhu, Lei Qian, Youling Yue, Zhichen Pan, Zhijie Liu, Xuhong Yu, Xiaoyao Xie, Qijun Zhi, Hui Zhang, Jumei Yao, Jun Yan, Chengmin Zhang, Paul S. Ray, Matthew Kerr, David A. Smith, Peter F. Michelson, Elizabeth C. Ferrara, David J. Thompson, Zhiqiang Shen, Na Wang, FAST\ --\ \textit{Fermi}-LAT Collaboration}

\address[{\rm1}]{CAS Key Laboratory of FAST, National Astronomical Observatories, Chinese Academy of Sciences, Beijing {\rm 100101}, China}
\address[{\rm2}]{University of Chinese Academy of Sciences, Beijing {\rm 100049}, China}     
\address[{\rm3}]{Jodrell Bank Centre for Astrophysics, Department of Physics and Astronomy, University of Manchester, Manchester M13 9PL, UK}   
\address[{\rm4}]{Department of Physics and Laboratory for Space Research, University of Hong Kong, Hong Kong {\rm 999077}, China}
\address[{\rm5}]{Santa Cruz Institute for Particle Physics, University of California, Santa Cruz, CA 95064, USA}
\address[{\rm6}]{Yunnan Observatories, Chinese Academy of Sciences, 396 Yangfangwang, Guandu District, Kunming {\rm 650216}, China}
\address[{\rm7}]{Key Laboratory for the Structure and Evolution of Celestial Objects, Chinese Academy of Sciences, 396 Yangfangwang, Guandu District, Kunming {\rm 650216}, China}
\address[{\rm8}]{School of Computer Science and Technology, Key Laboratory of Information and Computing Guizhou Province, Guizhou Normal University, Guiyang {\rm 550001}, China}
\address[{\rm9}]{School of Physics and Electronic Science, Guizhou Normal University, Guiyang {\rm 550001}, China}
\address[{\rm10}]{Phys. Dept., Stanford University, Stanford, California {\rm 94305}, USA}
\address[{\rm11}]{UMCP/CRESST/GSFC, NASA, Greenbelt, Maryland MD 20771, USA}
\address[{\rm12}]{GSFC, NASA, Greenbelt, Maryland MD 20771, USA}
\address[{\rm13}]{Centre d'\'Etudes Nucl\'eaires de Bordeaux Gradignan, IN2P3/CNRS, Universit\'e Bordeaux, 33175 Gradignan, France}
\address[{\rm14}]{Space Science Division, U.S. Naval Research Laboratory, Washington, DC 20375-5352 USA}
\address[{\rm15}]{Shanghai Astronomical Observatory, Chinese Academy of Sciences, Shanghai {\rm 200030}, China}
\address[{\rm16}]{Xinjiang Astronomical Observatory, 150, Science-1 Street, Urumqi, Xinjiang {\rm 830011}, China}
\address[{\rm17}]{Department of Physics, University of Maryland, College Park, MD 20742, USA}
\address[{\rm18}]{NAOC-UKZN Computational Astrophysics Centre, University of KwaZulu-Natal, Durban 4000, South Africa}

\abstract{High sensitivity radio searches of unassociated $\gamma$-ray sources have proven to be an effective way of finding new pulsars. Using the Five-hundred-meter Aperture Spherical radio Telescope (FAST) during its commissioning phase, we have carried out a number of targeted deep searches of \textit{Fermi} Large Area Telescope (LAT) $\gamma$-ray sources. On Feb. 27$^{th}$, 2018 we discovered an isolated millisecond pulsar (MSP), PSR J0318+0253, coincident with the unassociated $\gamma$-ray source 3FGL J0318.1+0252. PSR J0318+0253 has a spin period of $5.19$ milliseconds, a dispersion measure (DM) of $26$ pc cm$^{-3}$ corresponding to a DM distance of about $1.3$ kpc, and a period-averaged flux density of $\sim$11 $\pm$ 2 $\mu$Jy at L-band (1.05-1.45 GHz). Among all high energy MSPs, PSR J0318+0253 is the faintest ever detected in radio bands, by a factor of at least $\sim$4 in terms of L-band fluxes. With the aid of the radio ephemeris, an analysis of 9.6 years of \textit{Fermi}-LAT data revealed that PSR J0318+0253 also displays strong $\gamma$-ray pulsations. Follow-up observations carried out by both Arecibo and FAST suggest a likely spectral turn-over around 350 MHz.
This is the first result from the collaboration between FAST and the \textit{Fermi}-LAT teams as well as the first confirmed new MSP discovery by FAST, raising hopes for the detection of many more MSPs.  Such discoveries will make a significant contribution to our understanding of the neutron star zoo while potentially contributing to the future detection of gravitational waves, via pulsar timing array (PTA) experiments.}

\keywords{FAST, Pulsar, Radio, Gamma rays\\}

\PACS{97.60.Gb, 97.60.Jd, 95.85.Bh, 95.85.Pw\\}

\maketitle


\begin{multicols}{2}

\section{Introduction}\label{section1}
Millisecond pulsars (MSPs) are a special kind of old (or {\it recycled}) neutron star that rotates hundreds of times per second, emitting pulsations as their radiation beams sweep across our line of sight. MSPs are expected not only to play an important role in our understanding of the evolution of neutron stars and the equation of state of dense matter \cite{latimmer}, but can also be used  in Pulsar Timing Arrays (PTAs) \cite{Spolaor19}. PTAs attempt to detect low-frequency gravitational waves from merging supermassive black holes using the long-term timing of a set of stable MSPs with various radio telescopes around the globe. Pulsar searches to discover new MSPs with stable timing properties are essential to the ultimate goal of gravitational wave detection through PTAs and are also one of the primary science targets for the Five-hundred-meter Aperture Spherical radio Telescope (FAST) \cite{george} \cite{nan} \cite{li2016} \cite{li19}.

The \textit{Fermi} Large Area Telescope (LAT) has detected over 250 $\gamma$-ray pulsars \footnote{\url{https://confluence.slac.stanford.edu/display/GLAMCOG/Public+List+of+LAT-Detected+Gamma-Ray+Pulsars}} since its launch in 2008 \cite{atwood09}.  Roughly half of these are MSPs. Indeed, approximately one third of all known MSPs outside of globular clusters have been discovered by the Pulsar Search Consortium (PSC) \cite{psc}, a collaboration between the LAT team and radio astronomers, using the world's most sensitive radio telescopes to carry out follow-up blind searches of LAT sources. This effort has been tremendously successful, finding mostly {\it recycled} pulsars as well as a few {\it young} pulsars. Typically, non-varying, well-localized sources that are not associated with known Active Galactic Nuclei (AGN), perhaps with pulsar-like spectral characteristics, are targeted as potential pulsar candidates. In addition, because there is no empirical correlation between the radio and $\gamma$-ray fluxes, it is possible that faint LAT unassociated sources could yield just as many good radio pulsar candidates as the bright ones. This also emphasizes the importance of the high sensitivity of FAST in discovering faint MSPs in {\it pulsar-like} $\gamma$-ray sources that may have already been searched (unsuccessfully) by the other, less sensitive, radio telescopes in the PSC. The recent discovery of the first radio-quiet MSP \cite{J1035} establishes that a presently unknown population awaits characterization by instruments with FAST's sensitivity. A radio pulse can be faint or absent if the neutron star's magnetic and rotation inclinations relative to the Earth line-of-sight are such that the radio beam skims or misses Earth. Thus, a census of faint MSPs amounts to a rough mapping of MSP radio beams, imposing useful constraints on pulsar emission models.

In December 2017, a Memorandum of Understanding (MoU) was signed between the FAST team and the LAT Collaboration \cite{psc}, in order to share expertise and resources with the goal of making new pulsar discoveries. FAST is now transitioning from construction and commissioning to a fully operational Chinese National Facility \cite{jiang2019}. The ongoing surveys, such as the Search for Pulsars in Special Populations (SP$^2$ \cite{pan20}), the Commensal Radio Astronomy FAST Survey (CRAFTS \footnote{\url{http://crafts.bao.ac.cn/}} , \cite{li2018}, \cite{cameron20}) and the FAST Galactic Plane Pulsar Snapshot survey (GPPS\footnote{\url{http://zmtt.bao.ac.cn/GPPS/}}, \cite{han2021}) are expected to discover many more MSPs, and thus make a significant contribution to the PTA experiments. In this letter, we present the first MSP discovered by FAST, as part of the new joint FAST \ --\ \textit{Fermi}-LAT collaborative efforts.

\section{Source selection and observations}\label{sec:2}
The \textit{Fermi}-LAT Third Source Catalog (3FGL) contains 3033 $\gamma$-ray sources, including over 1,000 with no known counterparts at other wavelengths \cite{3FGL}. Using a number of statistical techniques, many of these so-called {\it unassociated} sources have been determined to be good pulsar candidates \cite{SazParkinson16}. In January 2018, the {\it Fermi}-LAT Collaboration released, via the {\it Fermi} Science Support Center (FSSC), a new list of over 5000 $\gamma$-ray sources (including $>$2000 {\it unassociated} sources) based on the analysis of 8 years of LAT data\footnote{\url{https://fermi.gsfc.nasa.gov/ssc/data/access/lat/fl8y/}}.
 We initially constructed our FAST target list based on the preliminary 8-year {\it Fermi} source list known as FL8Y, which was later superseded by the official \textit{Fermi}-LAT Fourth Source Catalog (4FGL) \cite{4FGL}. Following \cite{SazParkinson16}, we applied a number of machine learning algorithms to classify $\gamma$-ray sources into pulsar (PSR) and Active Galactic Nuclei (AGN) candidates. Specifically, we used Random Forests (RFs), logistic regression, Support Vector Machines (SVMs), and an Artificial Neural Network, to classify LAT unassociated sources into PSR and AGN candidates, based on their $\gamma$-ray features, most of which were provided in the publicly-available FL8Y list. Because the FL8Y list did not contain information on the variability of the $\gamma$-ray sources, we constructed a {\it variability index} for each source, based on its monthly variance in flux, computed using aperture photometry, following the procedure described on the FSSC web page\footnote{\url{https://fermi.gsfc.nasa.gov/ssc/data/analysis/scitools/aperture_photometry.html}}. For more details on the classification and ranking of FL8Y/4FGL $\gamma$-ray sources, see \footnote{P. M. Saz Parkinson et al. \ (2021), {\it in preparation}.}. Using two different samples of {\it identified} $\gamma$-ray sources to train our algorithms, one to train our algorithms and another to test them on, we obtained an overall accuracy of $>$95\%. We then picked targets from among LAT unassociated sources that were identified as likely PSRs by all four methods and selected those within the declination range accessible to FAST (from -14\degree to +66\degree). Of these, FL8Y J0318.2+0254 (aka 3FGL J0318.1+0252) was the {\it fourth} most significant $\gamma$-ray source (with a significance of 21.8 $\sigma$ in 8 years).

Early FAST commissioning observations were carried out in multiple observation modes, including pointing, drift scan, and tracking. A drift scan can be performed throughout the night, i.e. local time from 20:00 until 08:00 of the next day, so as to minimize uncertainties of the actuator system control, while the tracking and other non-static observations can be tested during the day.  These observations employ an actuator system (more than 2,100 actuators anchored to the ground will actively pull the connecting nodes and change the tension distribution of the cable-mesh system.) in which $\sim$1,000 points \cite{jiang2011} are measured and driven to achieve pointing and tracking, a substantially more complex system than conventional antennas which are driven along just two axes of motion, e.g., azimuth and elevation. A single beam, ultra-wide bandwidth receiver system in the frequency range of 270 MHz$-$1.62 GHz was installed in June of 2017, taking up a major portion of the receiver's commissioning time. The system temperature was 60$-$70 K across most of the band, the averaged efficiency was about 50$\%$ in the whole band of frequencies, and the gain was estimated to be 10.2 K/Jy during the observation. The half power beam width of $\sim$12 arcmin (300 MHz) covers the entire $\gamma$-ray position uncertainty region for most sources, making it convenient for performing a targeted deep integration. The low frequency system on FAST was decommissioned in April of 2018. Since then, the FAST L-band Array of 19 feed-horns (FLAN) \cite{li2018}, covering 1.05 - 1.45 GHz with a gain of about 16 K/Jy and a system temperature of about 20 K,  has been used for the major surveys.

The archived FAST data stream for pulsar observations is a time series of total power per frequency channel, stored in PSRFITS format \cite{hotan2004} from a ROACH-2 \footnote{\url{https://casper.ssl.berkeley.edu/wiki/ROACH-2_Revision_2}} based backend, which produces 8-bit sampled data over 4k frequency channels at 98 $\mu$s cadence.

\begin{figure}[H]
\centering
\includegraphics[scale=0.1]{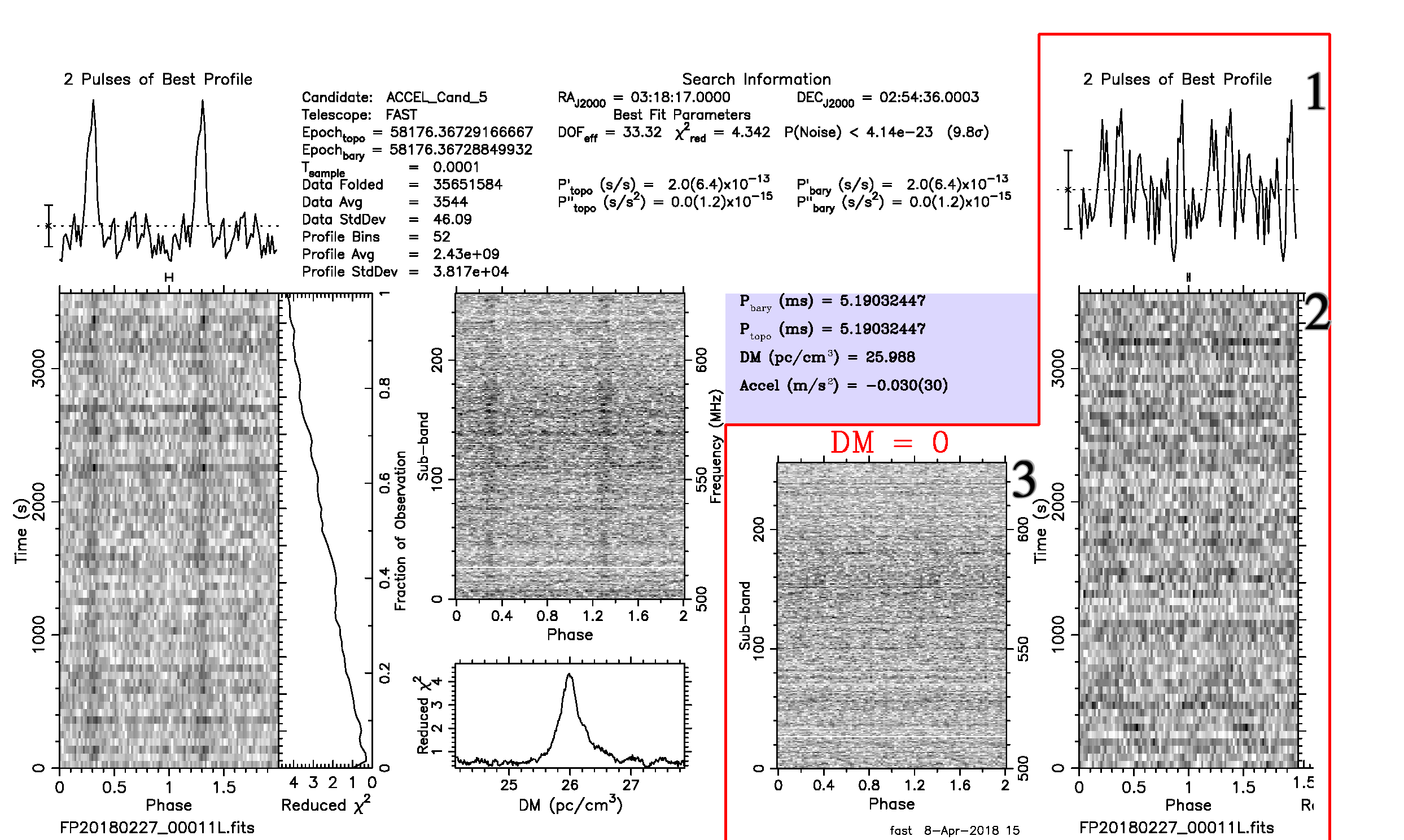}
\caption{The discovery diagnostic plot produced by the DPD pipeline for the 560-MHz candidate of PSR J0318+0253. Highlighted in the red box on the right, three subplots of DM=0 were added to the PRESTO output for better distinguishing signals from radio frequency interference (RFI). They are 1) integrated pulse profile; 2) intensities on phase vs.\ integration time plane; 3) intensities on phase vs.\ radio frequency. To the left of the red box are the candidate plots of PSR J0318+0253, adopted from PRESTO. The fold parameters are recorded in the purple box (highlighted on the left}.
\label{fig:fold}
\end{figure}

On 27$^{th}$ February 2018, in an one-hour tracking observation with the FAST ultra-wide band receiver, a promising  pulsar candidate was detected using the Distributed Presto with Database (DPD) pipeline (see the following section) toward the unassociated $\gamma$-ray source 3FGL J0318.1+0252.  We cut data into 20 minutes chunks and were able to identify the candidate in each. Previous radio observations of 3FGL J0318.1+0252, including three different epochs in June 2013 with Arecibo with typical individual integration times of 15 minutes, did not detect the MSP \cite{cromartie16}.

\section{Data analysis and result}\label{sec:3}
Largely based on the PRESTO package \cite{ransom}, we built a hybrid pulsar search pipeline, namely the Distributed Presto with Database (DPD), which sends packets of data with instructions to $\sim$ 100 PCs based on dynamic scheduling, utilizes our own GPU de-dispersion code,  incorporates DM=0 results into the candidate plots (Figure \ref{fig:fold}), and finally stores the ranked candidates in a database after sifting and AI ranking. DPD also utilizes a suite of machine-learning tools,  such as linear-feature enhancement, PICS \cite{zhu}, SPINN \cite{Morello}, and DCGAN \cite{guo}. More AI tools are being tested and can be easily incorporated into DPD. When needed, it is straightforward to use SQL queries to alter the combination of AI ranking and other selection criteria to filter the stored candidates in post-processing. 

PSR J0318+0253 was identified using DPD in a 512-MHz-wide band centered around 560 MHz. The refined parameters, including a DM = 26 pc cm$^{-3}$ and a spin period of 5.19 ms can be found in Table \ref{tab:J0318_info}. We estimated its 560-MHz flux density to be $\sim$100 $\mu$Jy using the radiometer equation \cite{lk2004} with $\sim$ 25$\%$ uncertainty. Within the uncertainty of the flux measurements and stability threshold, no flux variation was seen during the observation (1-2 hours integration time for each epoch), consistent with an apparent lack of scintillation.

Following the discovery of the radio pulsar, we performed a search for $\gamma$-ray pulsations using 9.6 years of \textit{Fermi}-LAT data \cite{atwood09}. We included ``Pass 8''  \cite{atwood13} SOURCE-class photons from within a 2$^{\circ}$ region around the center of the FL8Y\footnote{\url{https://fermi.gsfc.nasa.gov/ssc/data/access/lat/fl8y/}} region of interest (ROI), and with energies $100\,\textrm{MeV} < E < 30\,\textrm{GeV}$. To improve sensitivity, we computed a weight for each photon that represents its probability of having been emitted by the targeted source based on the photon's reconstructed energy and arrival direction, according to a model for the $\gamma$-ray flux\cite{Kerr2011}. As the spin frequency, spin-down rate, and sky location were not known precisely enough from the initial radio discovery to phase-fold the \textit{Fermi}-LAT data, we employed an efficient ``semi-coherent'' algorithm \cite{Methods2014} to search for $\gamma$-ray pulsations. We searched a small range around the pulsation frequency from the radio discovery, over spin-down rates from $0$ to $-10^{-14}\,\textrm{Hz s}^{-1}$, and over a conservatively wide sky region around the FL8Y source ROI. A highly significant signal was found. A fully-coherent refinement search around the candidate signal parameters found additional power in higher harmonics of the candidate frequency, for an eventual $H$-test \cite {dejager89} value of $H = 540$, corresponding to a single-trial false alarm probability of $\textrm{e}^{-0.398405\, H} \approx 10^{-93}$ \cite{Kerr2011}, confirming the detection.

Following this detection, the $\gamma$-ray spectrum of the source was refined using a binned likelihood analysis  \cite{Kerr2010}. The resulting spectral model was used to re-compute the photon probability weights, and a refined timing solution was obtained by performing an unbinned timing analysis using the $\gamma$-ray photon arrival times. The resulting $\gamma$-ray timing solution has a best-fitting position of RA(J2000)=03:18:15.541(1), Dec(J2000)=$+$02:53:01.48(5), with 1$\sigma$ uncertainties in the final digits given in parentheses, leading to the MSP being named as PSR J0318+0253. No significant proper motion was found, with a $95$\% confidence upper limit of $|\mu| < 37\,\textrm{mas yr}^{-1}$. No orbital motion was detected, confirming that it is an isolated MSP. The $\gamma$-ray timing analysis also results in a precise spin period of 5.19 milliseconds and spin-down rate measurement of $\dot{\nu} = -6.54(1)\times10^{-16}\,\textrm{Hz s}^{-1}$, corresponding to a spin-down power of $\dot{E} = -4 \pi^2 I \nu \dot{\nu} \approx5\times10^{33}\,\textrm{erg s}^{-1}$. The radio dispersion measure of 26 pc cm$^{-3}$ corresponds to a distance of about 1.3 kpc based on NE2001 \cite{ne2001} and YMW16 \cite{ymw16} galactic electron-density models. At the DM distance, the measured $\gamma$-ray energy flux corresponds to a luminosity of $\sim$1.2$\times10^{33}\,\textrm{erg s}^{-1}$, or approximately 25$\%$ of the total spin-down power. The resulting rotational and astrometric parameters on the timing are given in Table \ref{tab:J0318_info}.

\begin{figure}[H]
\centering
\includegraphics[scale=0.25]{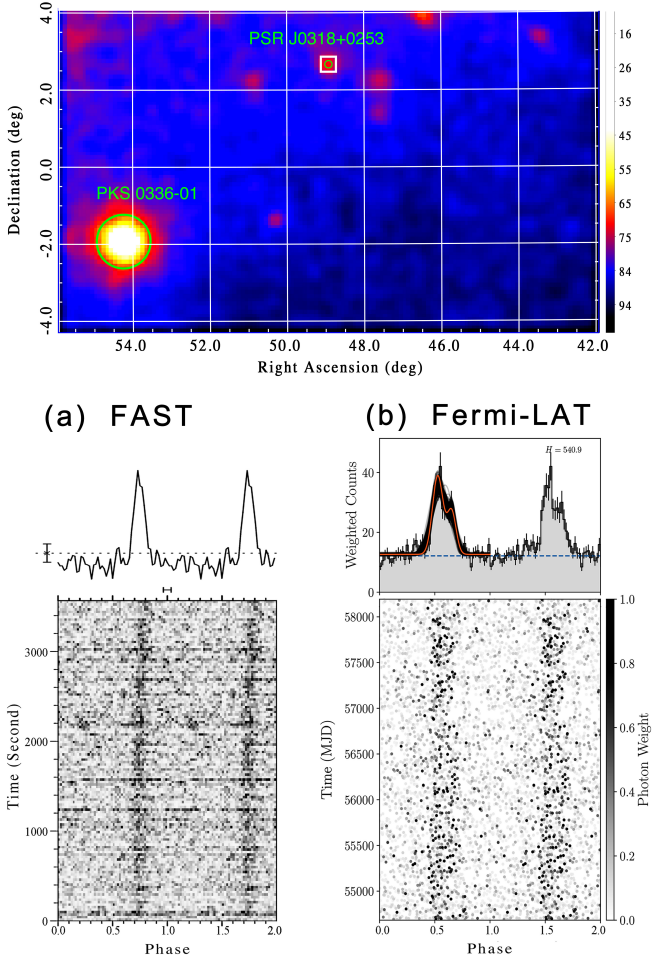}
\caption{The $\gamma$-ray sky map and integrated pulse profiles of the new MSP: Upper panel shows the region of the $\gamma$-ray sky where the new MSP is located. Lower panel (a) shows the observed radio pulses and integrated pulse profile in a one-hour tracking observation with FAST 560 MHz. Lower panel (b) shows the rotational phases and the integrated pulse profile of $\gamma$-ray photons observed from more than 9 years of \textit{Fermi}-LAT data.}
\label{fig:profile}
\end{figure}

The integrated pulse profiles of PSR J0318+0253 are shown in Figure \ref{fig:profile}, both with arbitrary phase alignment, as varying observational conditions during FAST instrument commissioning prevented us from obtaining the accurate clock corrections necessary to determine the radio reference epoch. We re-analysed X-ray archival data-sets and performed additional deeper radio observations in order to study the pulsar's multi-wavelength properties. The region around PSR J0318+0253 was observed for 3.5\,ks in the X-ray band, using the {\it Neil Gehrels Swift Observatory}, as part of a {\it Swift}-XRT survey of {\it Fermi}-LAT unassociated sources \cite{stroh2013}. No X-ray counterparts were detected at the pulsar position. By assuming a power-law spectrum of index of 2, and estimating\footnote{We use the $n_H$ tool available on the HEASARC web site: \url{https://heasarc.gsfc.nasa.gov/cgi-bin/Tools/w3nh/w3nh.pl}} the capitalize Galactic absorption column to be $n_H$ = $8\times10^{20}$ cm$^{2}$ in the direction of the MSP, we obtain an upper limit on the unabsorbed flux in the 0.3-10 keV energy range of $2.8\times10^{-13}$ erg cm$^{-2}$ s$^{-1}$. 

With the refined emphemeris, we carried out radio timing observations with the Arecibo Observatory at 327 MHz and with FAST in L-band. The pulsar was marginally visible in each FAST observation on 29th and 30th April 2019. We calibrated the noise level of the baseline, and then measured the amount of pulsed flux above the baseline, giving the flux measurement of 11 $\pm$ 2 $\mu$Jy (with 2 hours integration time at 1.25 GHz) for persistent radio pulsations.  Arecibo observations on 13th and 14th of July, 2019, resulted in non-detections at 327 MHz. The integration time of 1.2 hours in each epoch corresponds to a profile-averaged  5$\sigma$ upper limit of flux density at S (350 MHz) = $\sim$ 60 $\mu$Jy.

A power-law flux density function $S(\nu) \propto \nu^{\alpha}$ was fit to the flux densities measured by FAST at 560 MHz and 1.2 GHz, resulting in a spectral index of $\alpha = -1.9 \pm 0.2$. We can thus estimate the 350 MHz flux expected for Arecibo observations to be 130$\pm$30 $\mu$Jy, higher than our derived 5 sigma limit above. Although we cannot rule out diffractive and refractive scintillation effects from our limited number of observations, this discrepancy could indicate a turn-over in the spectrum around 350 MHz. A number of MSPs show evidence for spectral turn-over in the hundreds of MHz frequency range \cite{Kuzmin01} \cite{Kijak09} \cite{Kuniyoshi15} \cite{Kondratiev16} that could be due to the environmental conditions around neutron stars, or may be intrinsic to the radio emission mechanism. With the decommissioning of the ultra-wide-band receiver, FAST is yet to be equipped with more sensitive low-frequency ($<$1\,GHz) capabilities. Deeper P-band observations of PSR J0318+0253 are required to characterize its possible spectral turn-over and may shed light on the nature of this phenomenon.

The flux density of PSR J0318+0253 is measured to be $\sim$100 $\mu$Jy at 560 MHz and $\sim$11$\pm$2 $\mu$Jy at 1.2GHz with a spectral index of -1.9. Including this new discovery, there are 106 known high-energy MSPs (period $<$ 30 ms) seen with radio pulsations \footnote{ATNF Pulsar Catalogue: \url{https://www.atnf.csiro.au/research/pulsar/psrcat/}} \cite{psrcat}. Before PSR J0318+0253, the faintest radio pulses came from J1035-6720 with S$_{1400}$MHz $\sim$ 0.04 mJy \cite{J1035}, nearly 4 times brighter than the L-band flux of PSR J0318+0253. PSR J0318+0253 is thus the faintest high energy MSP discovered in the radio band to date.

\section{Conclusions}\label{sec:4}
We have performed targeted deep searches of {\it Fermi-LAT} unassociated $\gamma$-ray sources during the FAST commissioning phase. Using an augmented PRESTO-based pipeline, namely DPD, we discovered an isolated MSP coincident with the unassociated $\gamma$-ray source 3FGL J0318.1+0252. We later confirmed this discovery through the detection of $\gamma$-ray pulsations in the \textit{Fermi}-LAT data, now known as PSR J0318+0253, which represents the first discovery resulting from a collaboration between the FAST team and the \textit{Fermi}-LAT team. The pulsar has been added to the catalog of Galactic MSPs\footnote{\url{http://astro.phys.wvu.edu/GalacticMSPs/GalacticMSPs.txt}}. PSR J0318+0253 is the first MSP that FAST discovered and obtained full timing solution \footnote{\url{https://crafts.bao.ac.cn/pulsar/papers/fermi/}}, as part of the pilot program for FAST pulsar surveys, such as SP$^2$ \cite{pan20} and CRAFTS \cite{li2018}. 

Our multi-band analysis find that PSR J0318+0253 has a spin frequency of $192.68$ Hz (corresponding to a spin period of $5.19$ milliseconds), a spin-down power $\dot{E}$ of $5\times10^{33}$ erg s$^{-1}$, and a possible spectral turn-over at around 350 MHz. The 560 MHz flux density of $\sim$100 $\mu$Jy and the 1.2 GHz flux density of 11$\pm$2 $\mu$Jy make PSR J0318+0523 the faintest high energy MSP ever discovered in the radio bands. Pulsar searches with FAST, with the best absolute sensitivities at and below L-band, not only will result in numerous new pulsar discoveries, but will thus be important for complementing radio searches by other telescopes of high energy neutron stars and candidates to shed new lights into our inventory of neutron star population.

 \Acknowledgements{This work is supported by the National Natural Science Foundation of China  No. 11988101, 11690024, 11743002, 11873067, U1631132, U1831131, U1731238, U1938103, 11703047, 11773041, 11673060, National Key R\&D Program of China No. 2017YFA0402600, the Chinese Academy of Sciences (CAS) Key Laboratory of FAST, NAOC, Chinese Academy of Sciences, National Basic Research Program of China (973 program) No. 2015CB857100, the CAS Strategic Priority Research Program No. XDB23000000, the CAS International Partnership Program No. 114A11KYSB20160008, the Open Project Program of the Key Laboratory of FAST, NAOC, Chinese Academy of Sciences and Guizhou Provincial Key Laboratory of Radio Astronomy and Data Processing, Guizhou Normal University,Guiyang 550001,China. PW acknowledges support by the Youth Innovation Promotion Association CAS (id.~2021055), CAS Project for Young Scientists in Basic Reasearch (grant~YSBR-006) and the Cultivation Project for FAST Scientific Payoff and Research Achievement of CAMS-CAS, Z.P and X.H are supported by the CAS “Light of West China” Program. FAST is a Chinese national mega-science facility, built and operated by the National Astronomical Observatories, Chinese Academy of Sciences (NAOC). We appreciate all of the people in the FAST group for their support and assistance during the observations.
The \textit{Fermi}-LAT Collaboration acknowledges generous ongoing support
from a number of agencies and institutes that have supported both the
development and the operation of the LAT as well as scientific data analysis.
These include the National Aeronautics and Space Administration and the
Department of Energy in the United States, the Commissariat \`a l'Energie Atomique
and the Centre National de la Recherche Scientifique / Institut National de Physique
Nucl\'eaire et de Physique des Particules in France, the Agenzia Spaziale Italiana
and the Istituto Nazionale di Fisica Nucleare in Italy, the Ministry of Education,
Culture, Sports, Science and Technology (MEXT), High Energy Accelerator Research
Organization (KEK) and Japan Aerospace Exploration Agency (JAXA) in Japan, and
the K.~A.~Wallenberg Foundation, the Swedish Research Council and the
Swedish National Space Board in Sweden. Additional support for science analysis during the operations phase is gratefully acknowledged from the Istituto Nazionale di Astrofisica in Italy and the Centre National d'\'Etudes Spatiales in France. This work performed in part under DOE
Contract DE-AC02-76SF00515. This work was partially supported by the {\it Fermi} Guest Observer Program, administered by NASA under Grant No. 80NSSC18K1731. C.J.C acknowledges support from the ERC under the European Union's Horizon 2020 research and innovation 
programme (grant agreement No. 715051; Spiders). Fermi LAT work at NRL is supported by NASA.}



\def \apj {ApJ}
\def \apjl {ApJL}
\def \aap {A\&A}
\def \atel {The Astronomer's Telegram}
\def \apjs {ApJS}

\end{multicols}

\clearpage
\setlength{\tabcolsep}{1mm}{
\renewcommand\arraystretch{0.9}
\scriptsize
\begin{longtable}{c c c c c c c c c}%
\caption{Measured emission, rotational, and astrometric parameters of PSR J0318+0253.} \label{tab:J0318_info}
\\
\hline%
\hline%
\\
Spin frequency$^{a)}$&Spin frequency derivative$^{a)}$&Spin-down power$^{a)}$&Right ascension$^{a)}$&Declination$^{a)}$&DM&Flux Density&Spectral index$^{b)}$&Reference time \&\\%
$\nu$ (Hz)&$\dot{\nu}$ (Hz s$^{-1}$)&$\dot{E}$ (erg s$^{-1}$)&R.A.J2000 (h:m:s)&Dec.J2000 (d:m:s)&(pc cm$^{-3}$)&($\mu$Jy)&$\alpha$&Fermi Data span\\%
\\
\hline%
\endhead%
\hline%
\endfoot%
\hline%
\endlastfoot%
\\
&&&&&&S$_{350~MHz} < $ 60&&56437~(radio)\\%
192.68371552268(5)&$-6.54(1)\times10^{-16}$&$5\times10^{33}$&03:18:15.541(1) &$+$02:53:01.48(5) & 25.98(8) & S$_{560~MHz}  \sim$100 & -1.9(2) &  54681$-$58194 (9.6 yr)$^{a)}$\\%
&&&&&&S$_{1250~MHz}$ = 11(2)&&\\%
\\
\end{longtable}}
\begin{tablenotes}
\item[\#] $\#$ Values in parentheses represent 1$\sigma$ uncertainties on the last quoted digit.
\item[*] $*$ Clock correction procedure: TT(TAI), Solar system ephemeris model: DE421, Time unit: TCB.
\item[a)] $a)$ The rotational parameters are timed from $\gamma$-ray data.
\item[b)] $b)$ Only two of the fluxes (S$_{560~MHz}$ and S$_{1250~MHz}$) are used to compute the spectral index.
\end{tablenotes}


\begin{thebibliography}{99}
\bibitem{latimmer} J. M. Lattimer and M. Prakash, Physics Reports 333, 121 (2000).
\bibitem{Spolaor19} S. B. Spolaor, S. R. Taylor, and M. Charisi  et al., Astron Astrophys Rev, 27, 5 (2019).
\bibitem{george} G. Hobbs, S. Dai, and R. N. Manchester et al., the special issue on FAST in Research in Astronomy and Astrophysics (RAA), 019, 002 (2019).
\bibitem{nan} R. D. Nan, D. Li, C. J. Jin, Q. M. Wang, L. C. Zhu, W. B. Zhu, H. Y. Zhang, Y. L. Yue, and L. Qian, International Journal of Modern Physics D, 20, 989 (2011).
\bibitem{li2016} D. Li, and Z. Pan, Radio Science, 51, 1 (2016).
\bibitem{li19} D. Li, J. M. Dickey, and S. Liu, Research in Astronomy and Astrophysics, 19, 016 (2019).
\bibitem{atwood09} W. B. Atwood,  A. A. Abdo, and M. Ackermann et al., \apj, 697, 1071 (2009).
\bibitem{psc} P. S. Ray, A. A. Abdo, and D. Parent et al., arXiv:1205.3089 (2012).
\bibitem{jiang2019}  P. Jiang, Y. L. Yue, and H. Q. Gan et al., Sci. China-Phys. Mech. Astron. 62, 959502 (2019).
\bibitem{pan20} Z. Pan, S. M. Ransom, and D.R. Lorimer et al., \apjl, 892, L6 (2020).
\bibitem{li2018} D. Li, P. Wang, and L. Qian et al., IEEE Microwave Magazine 19, 112 (2018).
\bibitem{cameron20} A. Cameron, D. Li, and G. Hobbs et al., MNRAS, 95, 3515 (2020).
\bibitem{han2021} J. L. Han, C. Wang, and P. F. Wang et al., RAA, 21, 107 (2021).
\bibitem{3FGL} F. Acero, M. Ackermann, and M. Ajello et al., \apjs, 218, 23 (2015).
\bibitem{4FGL} S. Abdollahi, F. Acero, and M. Ackermann et al., \apjs, 247, 33A (2020).
\bibitem{SazParkinson16} P. M. Saz Parkinson, H. Xu, and P. L. H. Yu et al., \apj, 820, 8 (2016).
\bibitem{jiang2011} P. Jiang,  Q. Wang, and Q. Zhao, Applied Mechanics and Materials, 94, 979 (2011).
\bibitem{hotan2004} A. W. Hotan, W. van Straten, and R. N. Manchester, PASA 21, 302 (2004).
\bibitem{zhu} W. W. Zhu et al., \apj, 781, 117 (2014).
\bibitem{Morello} V. Morello, E. D. Barr, and M. Bailes et al., MNRAS, 443, 1651 (2014).
\bibitem{guo} P. Guo, F. Q. Duan, and P. Wang et al., MNRAS, 490, 5424 (2019).
\bibitem{ransom} S.~M.~Ransom, New Search Techniques for Binary Pulsars, PhD Thesis, Harvard University, Cambridge, (2001).
\bibitem{Andersen} B. C. Andersen, and S. Ransom, \apjl, 863, L13 (2018).
\bibitem{cromartie16} H. T. Cromartie, F. Camilo, and M. Kerr et al., \apj, 819, 34 (2016).
\bibitem{lk2004} D. R. Lorimer, and M. Kramer, Handbook of pulsar astronomy, ~Cambridge observing handbooks for research astronomers, Vol.~4.~Cambridge, UK: Cambridge University Press, 4 (2004).
\bibitem{atwood13} W., {Albert}, A., {Baldini}, L., {et~al.}, in Proceedings of the 4th Fermi Symposium, ed. T.~J. {Brandt}, N.~{Omodei}, \& C.~{Wilson-Hodge}, eConf C121028, 8, arXiv:1303.3514  (2012).
\bibitem{Kerr2011} M. Kerr, \apj, 732, 38 (2011).
\bibitem{Methods2014} H. J. Pletsch, and C. J. Clark., \apj, 795, 75  (2014).
\bibitem{dejager89} O. C. de Jager, B. C. Raubenheimer, and J. W. H. Swanepoel, \aap, 221, 180 (1989).
\bibitem{Kerr2010} M. Kerr, “Likelihood methods for the detection and characterization of gamma-ray pulsars with the Fermi Large Area Telescope,” thesis, University of Washington (2010).
\bibitem{ne2001} J. M. Cordes, and T. J. W. Lazio, arXiv:astro-ph/0207156 (2002).
\bibitem{ymw16} J. M. Yao, R. N. Manchester, and N. Wang, ApJ, 835, 29 (2017).
\bibitem{stroh2013} M. C. Stroh, and  A. D. Falcone, \apjs, 207, 28 (2013).
\bibitem{Kuzmin01} A. D. Kuzmin, and B. Ya. Losovsky, \aap, 368, 230 (2001).
\bibitem{Kijak09} J. Kijak, W. Lewandowski, and Y. Gupta, ASP Conference Series, Vol. 407 (2009).
\bibitem{Kuniyoshi15} M. Kuniyoshi, J. P. W. Verbiest, and K. J. Lee et al., MNRAS 453, 828 (2015).
\bibitem{Kondratiev16} V. I. Kondratiev, J. P. W. Verbiest, and J. W. T. Hessels et al., \aap, 585, A128 (2016).
\bibitem{J1035} C. J. Clark, H. J. Pletsch, J. Wu, L. Guillemot, M. Kerr, T. J. Johnson, F. Camilo, and D. Salvetti et al., Science Advances, 4, eaao7228 (2018).
\bibitem{psrcat} R. N. Manchester, G. Hobbs, A. Teoh, and M. Hobbs, AJ, 129, 1993 (2005).

\end{thebibliography}
\end{document}